\documentclass[a4paper]{article}

\usepackage{INTERSPEECH2018}
\usepackage{amsmath}
\usepackage[font=small,labelfont=bf]{caption}
\usepackage{graphicx}
\usepackage{capt-of}% or \usepackage{caption}
\usepackage{booktabs}
\usepackage{amsmath}
\usepackage[lowtilde]{url}

\def\newpara{\vspace{2pt}}
\def\aftertab{\vspace{-22pt}}
\def\subsec{\vspace{-5pt}}

\title{VoxCeleb2: Deep Speaker Recognition}
\name{Joon Son Chung$^\dag$, Arsha Nagrani$^\dag$, Andrew Zisserman
}
\address{
  Visual Geometry Group, Department of Engineering Science,\\
  University of Oxford, UK}
\email{\{joon,arsha,az\}@robots.ox.ac.uk}

\begin{document}

\maketitle
\begin{abstract}

The objective of this paper is speaker recognition under noisy and
unconstrained conditions.

We make two key contributions. First, we introduce a very large-scale~\textit{audio-visual} speaker recognition dataset collected from open-source media. Using a fully automated pipeline, we curate {\tt VoxCeleb2}
which contains over a million utterances from over 6,000 speakers. This is several times larger than any publicly available speaker recognition dataset.

Second, we develop and compare Convolutional Neural Network (CNN) models and training strategies that can effectively recognise identities from voice under various conditions.
The models trained on the  {\tt VoxCeleb2} dataset surpass the
performance of previous works on a benchmark dataset by a significant
margin. \\

\end{abstract}

\let\thefootnote\relax\footnotetext{\hspace{-12pt}$^\dag$These authors contributed equally to this work.}

%%  ---------- ---------- ---------- ---------- ---------- ---------- ---------- ----------
%%  ---------- ---------- ---------- ---------- ---------- ---------- ---------- ----------
\section{Introduction} 
Despite recent advances in the field of speaker recognition, producing single, compact representations for speaker segments that can be used efficiently under noisy and unconstrained conditions is still a significant challenge. In this paper, we present a deep CNN based neural speaker embedding system, named VGGVox, trained to map voice spectrograms to a compact Euclidean space where distances directly correspond to a measure of speaker similarity. Once such a space has been produced, other tasks such as speaker verification, clustering and diarisation can be straightforwardly implemented using standard techniques, with our embeddings as features.

Such a mapping has been learnt effectively for face images, through
the use of deep CNN
architectures~\cite{Schroff15,taigman2014deepface,Parkhi15} trained on
large-scale face
datasets~\cite{cao2017vggface2,kemelmacher2016megaface,guo2016ms}. Unfortunately,
speaker recognition still faces a dearth of large-scale freely
available datasets in the wild. \texttt{VoxCeleb1}~\cite{Nagrani17}
and SITW~\cite{Mclaren16} are valuable contributions, however they are
still an order of magnitude smaller than popular face datasets, which
contain millions of images. To address this issue we curate
\texttt{VoxCeleb2}, a large scale speaker recognition dataset obtained
automatically from open-source media. \texttt{VoxCeleb2} consists of
over a million utterances from over 6k speakers. Since the dataset is
collected `in the wild', the speech segments are corrupted with real
world noise including laughter, cross-talk, channel effects, music and
other sounds. The dataset is also multilingual, with speech from
speakers of 145 different nationalities, covering a wide range of
accents, ages, ethnicities and languages. 
% ** az: keep for the arxiv version
 The dataset is audio-visual,
 so is also useful for a number of other applications, for example --
 visual speech synthesis~\cite{Chung17b,karras2017audio}, speech separation~\cite{Afouras18,ephrat2018looking}, cross-modal transfer from face to voice or vice versa~\cite{Nagrani18a,nagrani2018learnable}
 and training face recognition
from video to complement existing face recognition datasets~\cite{cao2017vggface2,kemelmacher2016megaface,guo2016ms}. 
Both audio
 and video for the dataset will be released.

We train VGGVox on this dataset in order to learn speaker
discriminative embeddings. Our system consists of three main variable
parts: an underlying deep CNN~\textit{trunk} architecture, which is
used to extract the features, a pooling method which is used to
aggregate features to provide a single embedding for a given
utterance, and a pairwise loss trained on the features to directly
optimise the mapping itself. We experiment with both VGG-M~\cite{Chatfield14} and ResNet~\cite{He15} 
based trunk CNN architectures.

We make the following four contributions: (i) we curate and release a
large-scale dataset which is significantly larger than any other
speaker verification dataset. It also addresses a lack of ethnic
diversity in the \texttt{VoxCeleb1} dataset
(section~\ref{sec:dataset}); (ii) we propose deep ResNet-based
architectures for speaker embedding suitable for spectrogram inputs
(section~\ref{sec:vggvox}); (iii) we beat the current state of the art for speaker
verification on the \texttt{VoxCeleb1}  test set using our embeddings
(section~\ref{sec:results}); and (iv) we propose 
and evaluate on a new
verification benchmark test set which involves the entire \texttt{VoxCeleb1}  dataset.

\vspace{10pt}
The \texttt{VoxCeleb2} dataset
 can be downloaded from \url{http://www.robots.ox.ac.uk/~vgg/data/voxceleb2}.
 
%%  ---------- ---------- ---------- ---------- ---------- ---------- ---------- ----------
%%  ---------- ---------- ---------- ---------- ---------- ---------- ---------- ----------
\section{Related works}

\noindent{\bf Traditional methods.} 
Traditionally, the field of speaker recognition has been dominated by
i-vectors~\cite{dehak2011front}, classified using techniques such as
heavy-tailed PLDA~\cite{matvejka2011full} and
Gauss-PLDA~\cite{cumani2013probabilistic}. While defining  the
state-of-the-art for a long time, such methods are disadvantaged by
their reliance on hand-crafted feature engineering. An in-depth
review of these traditional methods is given in \cite{hansen2015speaker}.

\noindent{\bf Deep learning methods.}
The success of deep learning in computer vision and speech
recognition has motivated the use of deep neural networks (DNN) as
feature extractors combined with classifiers, though not trained end-to-end~\cite{variani2014deep,lei2014novel,ghalehjegh2015deep,snyder2017deep,snyder2018x}.
While such fusion methods are highly effective, they still require
hand-crafted engineering. In contrast, CNN architectures can be applied directly to raw
spectrograms and trained in an end-to-end manner. 
For example,
\cite{Chen11} uses a Siamese feedforward DNN to
discriminatively compare two voices, however this relies on
pre-computed MFCC features, whilst~\cite{Yella14} also learns the features instead of using MFCCs. 
The most relevant to our work is~\cite{li2017deep}, who train a neural embedding system using the triplet loss. However, they use private internal datasets for both training and evaluation, and hence a direct comparison with their work is not possible.

\noindent{\bf Datasets.}
Existing speaker recognition datasets usually suffer from one or more
of the following limitations: (i) they are either obtained under
controlled conditions (e.g.,  from telephone
calls~\cite{vannfi,Hennebert00} or acoustic
laboratories~\cite{Millar94,Garofolo93,Fisher86}), (ii) they are
manually annotated and hence limited in size~\cite{Mclaren16}, or
(iii) not freely available to the speaker
community~\cite{greenberg2012nist,Fisher86}  (see~\cite{Nagrani17} for
a full review of existing datasets). In contrast, the
\texttt{VoxCeleb2} dataset does not suffer from any of these limitations.

%%  ---------- ---------- ---------- ---------- ---------- ---------- ---------- ----------
%%  ---------- ---------- ---------- ---------- ---------- ---------- ---------- ----------
\section{The {\tt VoxCeleb2} Dataset}
\label{sec:dataset}

\subsection{Description}
\begin{figure*}[ht]
\vspace{-30pt}
\centering
\includegraphics[width=1\linewidth]{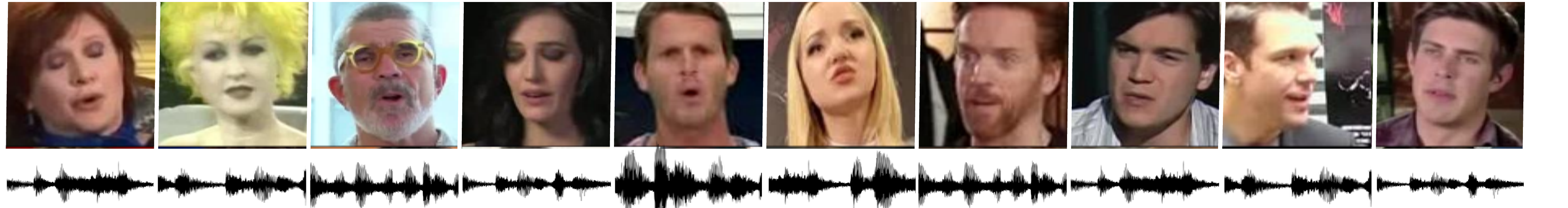}
\includegraphics[width=1\linewidth]{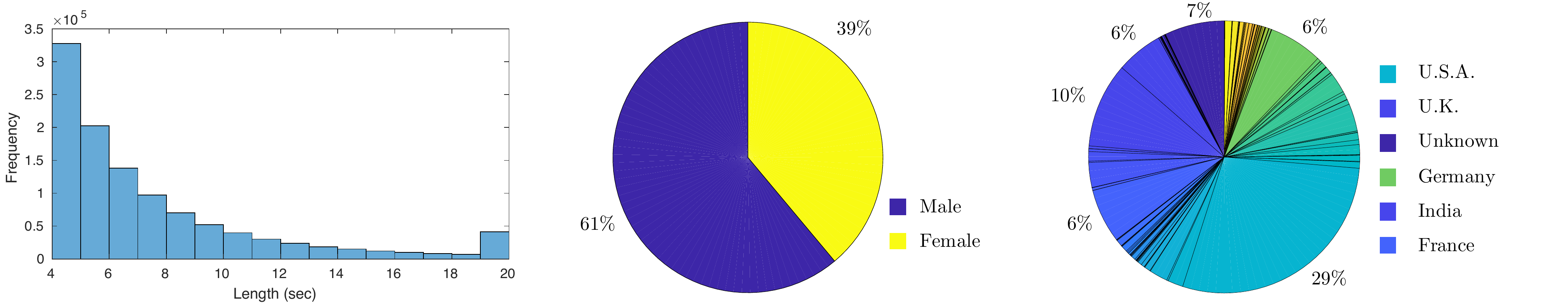}
   \caption{{\bf Top row}: Examples from the VoxCeleb2 dataset. We show cropped faces of some of the speakers in the dataset. Both audio and face detections are provided. {\bf Bottom row}: (left) distribution of utterance lengths in the dataset -- lengths shorter than $20s$ are binned in $1s$ intervals and all utterances of $20s+$ are binned together; (middle) gender distribution and (right) nationality distribution of speakers. For readability, the percentage frequencies of only the top-5 nationalities are shown. Best viewed zoomed in and in colour. }
\label{fig:vox2stats}
\vspace{-14pt}
\end{figure*}

\subsec
{\tt VoxCeleb2} contains over 1 million utterances for over 6,000
celebrities, extracted from videos uploaded to YouTube. The dataset is
fairly gender balanced, with 61\% of the speakers male. The speakers span a
wide range of different ethnicities, accents, professions and ages. 
Videos included in the dataset are shot in a large number of challenging visual and auditory environments. These include interviews from red carpets, outdoor stadiums and quiet indoor studios, speeches given to large audiences, excerpts from professionally shot multimedia, and even crude videos shot on hand-held devices. Audio segments present in the dataset are degraded with background chatter, laughter, overlapping speech and varying room acoustics. 
%while face images span a wide range of lighting conditions, resolution quality and pose variation. 
We also provide face detections and face-tracks for the speakers in the dataset, and the face images are similarly `in the wild', with variations in pose (including profiles), lighting, image quality and motion blur.  Table~\ref{table:data_stats} gives the general statistics, and Figure~\ref{fig:vox2stats} shows examples of cropped faces as well as utterance length, gender and nationality distributions. 

The dataset contains both development (train/val) and test sets.
However, since we use the {\tt VoxCeleb1} dataset for testing,  
only the development set will be used for the speaker
recognition task (Sections~\ref{sec:vggvox} and \ref{sec:results}). The VoxCeleb2 test set
should prove useful for other applications of audio-visual learning
for which the dataset might be used. The split is given in
Table~\ref{table:set_stats}.  The development set of {\tt VoxCeleb2}
has no overlap with the identities in the {\tt VoxCeleb1} or SITW
datasets.

\begin{table}
\footnotesize
\centering
\begin{tabular}{ l  r  r  }
  \textbf{Dataset} &  \textbf{VoxCeleb1} &  \textbf{VoxCeleb2} \\ \hline  
 \# of POIs & 1,251 & 6,112 \\ 
  \#  of male POIs & 690 & 3,761  \\ 
    \#  of videos & 22,496 & 150,480\\ 
    \#  of hours & 352 & 2,442\\ 
\#  of utterances & 153,516  & 1,128,246 \\ 
Avg \# of videos per POI & 18 & 25  \\ 
Avg \# of utterances per POI& 116 & 185 \\ 
Avg length of utterances (s) & 8.2 & 7.8\\ 

\end{tabular} 
\vspace{3pt}

\caption{Dataset statistics for both {\tt VoxCeleb1} and {\tt VoxCeleb2}. Note {\tt VoxCeleb2} is more than 5 times larger than {\tt VoxCeleb1}. POI: Person  of Interest.}
\label{table:data_stats}
\normalsize
\aftertab
\vspace{7pt}
\end{table}

\begin{table}
\footnotesize
\centering
\begin{tabular}{ l  r  r  r }

  \textbf{Dataset} &  \textbf{Dev} &  \textbf{Test} &  \textbf{Total}  \\ \hline  
 \# of POIs & 5,994 & 118 & 6,112 \\ 
    \#  of videos & 145,569 & 4,911 & 150,480\\ 
\#  of utterances & 1,092,009 & 36,237 & 1,128,246 \\ 

\end{tabular} 
\vspace{3pt}

\caption{Development and test set split.}
\label{table:set_stats}
\normalsize
\aftertab
\vspace{-8pt}
\end{table}

\subsection{Collection Pipeline}
\subsec

We use an automatic computer vision pipeline to
curate~\texttt{VoxCeleb2}. While the pipeline is similar to that used
to compile \texttt{VoxCeleb1}~\cite{Nagrani17}, the details have been
modified to increase efficiency and allow talking faces to be
recognised from multiple poses, not only near-frontal. In fact, we change the
implementation of every key component of the pipeline: the face
detector, the face tracker, the SyncNet model used to perform active
speaker verification, and the final face recognition model at the
end. We also add an additional step for automatic duplicate
removal. This pipeline allows us to obtain a dataset that is five times the size
of~\cite{Nagrani17}. We also note that the list of celebrity names
spans a wider range of nationalities, and hence 
unlike~\cite{Nagrani17}, the dataset obtained is multi-lingual.  For the
sake of clarity, the key stages are discussed in the following
paragraphs:

\newpara\noindent\textbf{Stage 1.\  Candidate list of Persons of Interest (POIs).}
The first stage is to obtain a list of POIs. We start from the list of people that appear in the VGGFace2 dataset~\cite{cao2017vggface2}, which has considerable ethnic diversity and diversity in profession. This list contains over 9,000 identities, ranging from actors and sportspeople to politicians. 
Identities that overlap with those of \texttt{VoxCeleb1} and SITW are removed from the development set. 

\newpara\noindent\textbf{Stage 2.\  Downloading videos.}
The top 100 videos for each of the POIs are automatically downloaded using YouTube search. The word `interview' is appended to the name of the POI in search queries to increase the likelihood that the videos contain an instance of the POI speaking, as opposed to sports or music videos. 

\newpara\noindent\textbf{Stage 3.\  Face tracking.}
The CNN face detector based on the
Single Shot MultiBox Detector (SSD)~\cite{Liu16}
is used to detect face appearances
 on every frame of the video. This detector is a distinct improvement  from that used in~\cite{Nagrani17}, allowing the detection of faces in profile and extreme poses. 
 We used the same tracker as~\cite{Nagrani17} based on ROI overlap.
 
\newpara\noindent\textbf{Stage 4.\  Face verification. }
A face recognition CNN is used to classify the face tracks into whether they are of the POI or not.
The classification network used here is based on the ResNet-50~\cite{He15} trained on the VGGFace2
dataset. Verification is done by directly using this
classification score. 

\newpara\noindent\textbf{Stage 5.\  Active speaker verification.}
The goal of this stage is to determine if the  visible face is the speaker. This is done by using a multi-view adaptation~\cite{Chung17a} of `SyncNet'~\cite{Chung16a,Chung18}, a two-stream CNN which determines the active speaker by estimating the correlation between the audio track and the mouth motion of the video. The method can reject clips that contain dubbing or voice-over. 

\newpara\noindent\textbf{Stage 6.\  Duplicate removal. } 
A caveat of using YouTube as a source for videos is that often the
same video (or a section of a video) can be uploaded twice, albeit
with different URLs. Duplicates are identified and removed as follows:
each speech segment  is represented by a  1024D  vector using
the model in~\cite{Nagrani17} as a feature extractor.
The Euclidean distance is computed between all pairs of 
features from the same speaker. If any two speech segments  have a
distance smaller than a very conservative threshold (of $0.1$), then the 
the speech segments are deemed to be identical, and one is removed. This method will certainly
identify all exact duplicates, and in practice we find that it also succeeds in identifying near-duplicates,
e.g.\ speech segments of the same source that are differently trimmed. 

\newpara\noindent\textbf{Stage 7.\  Obtaining nationality labels. } 
Nationality labels are crawled from Wikipedia for all the celebrities
in the dataset. We crawl for country of~\textit{citizenship}, and
not~\textit{ethnicity}, as this is often more indicative of accent. In
total,  nationality labels are obtained for all but 428 speakers, who
were labelled as unknown. Speakers in the dataset were found to hail
from $145$ nationalities (compared to $36$ for { \tt VoxCeleb1}),
yielding a far more ethnically diverse dataset (See
Figure~\ref{fig:vox2stats} (bottom, right) for the distribution of
nationalities). Note also the percentage of U.S. speakers is smaller in
{ \tt VoxCeleb2} ($29\%$) compared to { \tt VoxCeleb1} ($64\%$) where
it dominates. 

\newpara\noindent\textbf{Discussion. }  
In order to ensure that our system is extremely confident that a
person has been correctly
identified (Stage 4), and that are 
speaking 
 (Stage 5) without any manual interference, we set 
conservative thresholds in order to minimise the
number of false positives.  
Since \texttt{VoxCeleb2} is designed primarily as a training-only dataset,
the thresholds are less strict compared to those used to compile
\texttt{VoxCeleb1}, so that fewer videos are discarded.
Despite this, we have only found very few label errors after manual inspection
of a significant subset of the dataset.
%%  ---------- ---------- ---------- ---------- ---------- ---------- ---------- ----------
%%  ---------- ---------- ---------- ---------- ---------- ---------- ---------- ----------

%%  ---------- ---------- ---------- ---------- ---------- ---------- ---------- ----------
%%  ---------- ---------- ---------- ---------- ---------- ---------- ---------- ----------
\section{VGGVox}
\label{sec:vggvox}

In this section we describe our neural embedding system, called VGGVox.
  The system is trained on short-term magnitude spectrograms extracted directly from raw audio segments, with no other pre-processing. A deep
neural network trunk architecture is used to extract frame level
features, which are pooled to obtain utterance-level speaker
embeddings. The entire model is then trained using contrastive loss.
Pre-training using a soft-max layer and cross-entropy over a
fixed list of speakers improves model performance; hence we pre-train
the trunk architecture model for the task of identification first.

\subsection{Evaluation}
\subsec
\
The model is trained on the \texttt{VoxCeleb2} dataset.
At train time, pairs are sampled on-line using the method described in 
Section~\ref{sec:trainingloss}. The testing is done on the
\texttt{VoxCeleb1} dataset, with the test pairs provided
in that dataset.

We report two performance metrics:
(i) the Equal Error Rate (EER) which is the rate at 
which both acceptance and rejection errors are equal;
and
(ii) the cost function 
\begin{equation}
  C_{det} = C_{miss} \times P_{miss} \times P_{tar} + C_{fa} \times P_{fa} \times (1-P_{tar})
  \label{equ:cdet}
\end{equation}
where we assume a prior target probability $P_{tar}$ of 0.01 and
equal weights of 1.0 between misses $C_{miss}$ and false alarms $C_{fa}$.
Both metrics are commonly used for evaluating  identity 
verification systems.

\subsection{Trunk architectures}
\subsec

\newpara\noindent\textbf{VGG-M:}  The baseline trunk architecture is the CNN introduced
in~\cite{Nagrani17}. This architecture is a modification of
the VGG-M~\cite{Chatfield14} CNN, known for high efficiency and good
classification performance on image data.
In particular, 
the fully connected {\em fc6} layer from the original VGG-M is replaced by two layers -- a fully connected layer of $9 \times 1$ (support in the frequency domain), and an average pool layer with support $1 \times n$, where $n$ depends on the length of the input speech segment
(for example for a 3 second segment, $n=8$). 
The benefit of this modification is that the network becomes invariant
to temporal position but \textit{not} frequency, which is desirable for speech, but not for images. It also helps to keep the output dimensions the same as those of the original fully connected layer, and reduces the number of network parameters by fivefold.

\newpara\noindent\textbf{ResNets:} 
% While deeper networks have shown to improve performance for numerous
% classification tasks~\cite{Chatfield11}, they are notoriously
% difficult to train~\cite{He15}, with performance saturating
% after a certain number of layers. Hence deep residual
% networks~\cite{He15} were proposed to ease the training of very
% deep CNNs. 
The residual-network (ResNet) architecture~\cite{He15}  is similar to a standard
multi-layer CNN, but with added skip connections such that the layers add residuals to
an identity mapping on the channel outputs.
We experiment with both
ResNet-34 and ResNet-50 architectures, 
%which provides a good trade-off between
%computational efficiency and reasonable performance on image
%classification tasks, 
and modify the layers to adapt to the
spectrogram input. We apply batch normalisation before computing
rectified linear unit (ReLU) activations.  
The architectures are specified in Table~\ref{table:convnet}.

\begin{table}[ht]
\footnotesize
\centering
\begin{tabular}{ c|c|c }
\textbf{layer name} & \textbf{res-34}  & \textbf{res-50}   \\ \hline
conv1 & $ 7 \times 7,64$, stride 2 & $ 7 \times 7,64$, stride 2\\ 
%\addlinespace[0.5ex]
pool1 &  $ 3 \times 3$, max pool, stride 2 &  $ 3 \times 3$, max pool, stride 2 \\  \hline
%\addlinespace[1.5ex]
conv2\_x & $\begin{bmatrix}  3 \times 3, 64 \\ 3 \times 3, 64 \end{bmatrix} \times 3 $   & $\begin{bmatrix}  1 \times 1, 64 \\ 3 \times 3, 64 \\  1 \times 1, 256 \end{bmatrix} \times 3 $          \\\hline
%\addlinespace[1.5ex]
conv3\_x &$\begin{bmatrix}  3 \times 3,128 \\ 3 \times 3,128 \end{bmatrix} \times 4 $  & $\begin{bmatrix}  1 \times 1, 128 \\ 3 \times 3, 128 \\  1 \times 1, 512\end{bmatrix} \times 4 $           \\\hline
%\addlinespace[1.5ex]
conv4\_x & $\begin{bmatrix}  3 \times 3,256 \\ 3 \times 3,256 \end{bmatrix} \times 6 $   & $\begin{bmatrix}  1 \times 1, 256 \\  3 \times 3, 256 \\  1 \times 1, 1024 \end{bmatrix} \times 6 $          \\\hline
%\addlinespace[1.5ex]
conv5\_x &$\begin{bmatrix}  3 \times 3,512 \\ 3 \times 3,512 \end{bmatrix} \times 3 $     & $\begin{bmatrix}  1 \times 1, 512 \\ 3 \times 3, 512 \\ 1 \times 1, 2048 \end{bmatrix} \times 3 $            \\\hline
%\addlinespace[1.5ex]
fc1 &  $ 9 \times 1$, 512, stride 1  &  $ 9 \times 1$, 2048, stride 1                  \\ 
%\addlinespace[0.5ex]
pool\_time &  $ 1 \times N$, avg pool, stride 1  &  $ 1 \times N$, avg pool, stride 1\\ 
%\addlinespace[0.5ex]
fc2 &  $ 1 \times 1$, 5994 &  $ 1 \times 1$, 5994 \\ 
\end{tabular}
\vspace{2pt}
\normalsize
\caption{Modified Res-34 and Res-50 architectures with average pool layer at the end. ReLU and batchnorm layers are not shown.  Each row specifies the number of convolutional filters and their sizes as {\bf size $\times$ size, \# filters}.}
\label{table:convnet}
\aftertab
\vspace{-8pt}
\end{table}

\subsection{Training Loss strategies}
\subsec
\label{sec:trainingloss}
We employ a contrastive loss~\cite{Chopra05,hadsell2006dimensionality}
on paired embeddings, which seeks to minimise the distance between the
embeddings of positive pairs and penalises the negative pair distances
for being smaller than a margin parameter $\alpha$. Pair-wise losses
such as the contrastive loss are notoriously difficult to
train~\cite{hermans2017defense}, and hence to avoid suboptimal local
minima early on in training, we proceed in two stages:
first, pre-training for identifcation using a softmax loss, then,
second, fine-tuning with the contrastive loss.

\newpara\noindent\textbf{Pre-training for identification:} Our first strategy is to use  softmax pre-training to initialise the weights of the network. The cross entropy
loss produces more stable convergence than the contrastive loss, possibly because softmax training is not impacted by the 
difficulty of pairs when using the contrastive loss. To evaluate the identification performance, we create a held-out validation test which consists of all the speech segments from a single video for each identity. 

\newpara\noindent\textbf{Learning an embedding with contrastive loss -- hard negative mining:} 
We take the model pre-trained on the identification task, and replace the 5994-way classification layer with a fully connected layer of output dimension 512. This network is trained with contrastive loss.

A key challenge associated with learning embeddings via the contrastive loss is that as the dataset gets larger, the number of possible pairs grows quadratically. In such a scenario, the network rapidly learns to correctly map the easy examples, and hard negative mining  is often required to improve performance to provide the network with a more useful learning signal. We use an offline hard negative mining strategy, which allows us to select harder negatives ({\em e.g.} top 1-percent of randomly generated pairs) than is possible with online (in-batch) hard negative mining methods~\cite{sung1996learning,hermans2017defense,song2016deep} limited by the batch size. We do not mine hard positives, since false positive pairs are much more likely to occur than false negative pairs in a random sample (due to possible label noise on the face verification), and these label errors will lead to poor learning dynamics.

\subsection{Test time augmentation}
\subsec
\label{subsec:testaug}
\cite{Nagrani17} use average pooling at test time by evaluating the entire test utterance at once by changing the size of the {\it apool6} layer. Here, we experiment with different augmentation protocols for evaluating the performance at test time. We propose three methods:\\
\noindent{\bf (1)} Baseline: variable average pooling as described in~\cite{Nagrani17};\\
\noindent{\bf (2)} Sample ten 3-second temporal crops from each test segment, and use the mean of the features;\\
\noindent{\bf (3)} Sample ten 3-second temporal crops from each test segment, compute the distances between the every possible pair of crops ($10\times10=100$) from the two speech segments, and use the mean of the 100 distances.
The method results in a marginal improvement in performance, as shown in Table~\ref{table:results}.

\subsection{Implementation Details} 
\subsec
\newpara\noindent\textbf{Input features.} 
Spectrograms are computed from raw audio  in a sliding window fashion using a hamming window of width 25ms and step
10ms, in exactly the same manner as~\cite{Nagrani17}. This gives spectrograms of size 512 x 300 for
3 seconds of speech.  Mean and variance normalisation is performed on every frequency bin of the spectrum.

\newpara\noindent\textbf{Training.}
During training, we randomly sample 3-second segments from each utterance.
Our implementation is based on the deep learning toolbox MatConvNet~\cite{Vedaldi14a}.
Each network is trained on three Titan X GPUs for $30$ epochs or until the validation error stops decreasing, whichever is sooner, using a batch-size of $64$. We use SGD with momentum ($0.9$), weight decay ($5E-4$) and a logarithmically decaying learning rate (initialised to $10^{-2}$ and decaying to $10^{-8}$).

%%  ---------- ---------- ---------- ---------- ---------- ---------- ---------- ----------
%%  ---------- ---------- ---------- ---------- ---------- ---------- ---------- ----------
\section{Results}
\label{sec:results}
\newpara\noindent\textbf{Original \texttt{VoxCeleb1}  test set.}
Table~\ref{table:results} provides the performance of our models on
the original \texttt{VoxCeleb1} test set. As might be expected, performance improves with greater network depth, and
also with more training data (\texttt{VoxCeleb2} vs \texttt{VoxCeleb1}). This also demonstrates that 
\texttt{VoxCeleb2} provides a suitable training regime for use on other datasets.

\begin{table}[h!]
\centering
\footnotesize
\begin{tabular}{ l  l c r  r  }

\textbf{Models}             					& \textbf{Trained on} & $C^{min}_{det}$ & EER (\%) \\ \hline  
 I-vectors + PLDA (1)~\cite{Nagrani17}				& \texttt{VoxCeleb1}      & 0.73     & 8.8    \\
 VGG-M (Softmax)~\cite{Nagrani17}		& \texttt{VoxCeleb1} & 0.75    & 10.2 \\ 
 VGG-M 	(1)		~\cite{Nagrani17}		& \texttt{VoxCeleb1} & 0.71    & 7.8 \\ 
 VGG-M 	(1)									  	& \texttt{VoxCeleb2} & 0.609 & 5.94 \\  \hline  
 
 ResNet-34 	(1)							& \texttt{VoxCeleb2} &  0.543 & 5.04 \\
 ResNet-34 	(2)							& \texttt{VoxCeleb2} &  0.553 & 5.11 \\
 ResNet-34 	(3)							& \texttt{VoxCeleb2} &  0.549 & 4.83 \\ \hline  
 
 ResNet-50 	(1)							& \texttt{VoxCeleb2} & 0.449 & 4.19 \\
 ResNet-50 	(2)							& \texttt{VoxCeleb2} & 0.454 & 4.43 \\
 ResNet-50	(3)							& \texttt{VoxCeleb2} &  \textbf{0.429} & \textbf{3.95}  \\ \hline  

\end{tabular} 
\caption{Results for verification on the original {\tt VoxCeleb1} test set (lower is better). The number in brackets refer to the test time augmentation methods described in Section~\ref{subsec:testaug}.}
\label{table:results}
\normalsize
\aftertab
\end{table}

\newpara\noindent\textbf{New \texttt{VoxCeleb1-E} test set --  using the entire dataset.}
Popular speaker verification test sets in the wild~\cite{Nagrani17, Mclaren16} are limited in the number of speakers. This yields the possible danger of optimising performance to overfit
the small number of speakers in the test set, and results are not
always indicative of good generalised performance. Hence we propose a
new evaluation protocol consisting of 581,480 random pairs sampled from the
entire \texttt{VoxCeleb1} dataset, covering 1,251 speakers, and set benchmark
performance for this test set.
The result is given in Table~\ref{table:results_vox1e}.

\newpara\noindent\textbf{New \texttt{VoxCeleb1-H} test set --  within the same nationality and gender.}
By using the whole of \texttt{VoxCeleb1} as a test set,  we are able to test only on the pairs with same nationality and gender. 
We propose a new evaluation list consisting of 552,536 pairs sampled from the 
\texttt{VoxCeleb1} dataset, all of which are from the same nationality
and gender. 18 nationality-gender combinations each with at least 5 individuals are used
to generate this list, of which `USA-Male' is the most common.
The result is given in Table~\ref{table:results_vox1e}.

\begin{table}[h!]
\centering
\footnotesize
\begin{tabular}{ l l c r  r  }

\textbf{Models} & \textbf{Tested on} & $C^{min}_{det}$ & EER (\%) \\ \hline  
 ResNet-50 (3) 			& \texttt{VoxCeleb1-E} & {\bf 0.524} & {\bf 4.42} \\ 
 ResNet-50 (3) 			& \texttt{VoxCeleb1-H} & {\bf 0.673} & {\bf 7.33} \\ 
\end{tabular} 
\vspace{3pt}

\caption{Results for verification on the extended {\tt VoxCeleb1} test sets.}
\label{table:results_vox1e}
\normalsize
\aftertab
\vspace{-10pt}
\end{table}

%%  ---------- ---------- ---------- ---------- ---------- ---------- ---------- ----------
%%  ---------- ---------- ---------- ---------- ---------- ---------- ---------- ----------
\section{Conclusion}

In this paper, we have introduced new architectures and training strategies for the task of speaker verification, and demonstrated state-of-the-art performance on the \texttt{VoxCeleb1} dataset. 
%Since the improved hard negative mining method results in a significant boost to performance, a natural extension to this work would be to consider further changes to sampling strategies, such as hard positive mining. 
Our learnt identity embeddings are compact ($512D$) and hence easy to store and useful for other tasks such as diarisation and retrieval.
We have also introduced the \texttt{VoxCeleb2} dataset, which is several times larger than any speaker recognition dataset, and
have re-purposed the \texttt{VoxCeleb1}  dataset, so that the entire dataset 
of 1,251 speakers can be used as
a test set for speaker verification. Choosing pairs from all speakers allows a 
better assessment of performance than from the $40$ speakers of the original test set. We hope that this
new test set will be adopted, alongside SITW, as a standard for the speech community to
evaluate on.

 \newpara\noindent\textbf{Acknowledgements.}
 Funding for this research is provided by the EPSRC 
 Programme Grant Seebibyte EP/M013774/1. 

%\clearpage %

\bibliographystyle{IEEEtran}
\bibliography{shortstrings,vgg_other,mybib,vgg_local}

\end{document}